\documentclass[journal]{IEEEtran}
\usepackage{amsmath,amsfonts}
\usepackage{algorithmic}
\usepackage{algorithm}
\usepackage{xcolor}
\usepackage{array}
\usepackage[caption=false,font=normalsize,labelfont=sf,textfont=sf]{subfig}
\usepackage{textcomp}
\usepackage{stfloats}
\usepackage{url}
\usepackage{verbatim}
\usepackage{graphicx}
\usepackage{cite}
\usepackage[hidelinks]{hyperref}
\usepackage{float}
\usepackage{subcaption}

\begin{document}

\title{Joint Optimization of RU~Allocation and C-SR in Multi-AP Coordinated Wi-Fi Systems}

\author{Md Rahat Hasan,
        Kazi Ahmed Akbar Munim,
        and Md. Forkan Uddin, \IEEEmembership{Senior Member,~IEEE}
      
\thanks{Md Rahat Hasan, Kazi Ahmed Akbar Munim, and Md. Forkan Uddin are with the Department~of~Electrical and Electronic Engineering, Bangladesh~University~of~Engineering~and~Technology, Dhaka, Bangladesh (e-mail:~mforkanuddin@eee.buet.ac.bd).}
\thanks{Md. Rahat Hasan and Kazi Ahmed Akbar Munim have contributed to this work equally.}
 
}



\maketitle

\begin{abstract}
We formulate an optimization problem for joint RU allocation and C-SR to maximize the throughput of a multi-AP coordinated WiFi system. The optimization problem is found to be a non-linear integer programming problem. We solve the problem for several network scenarios using an optimization tool. The joint design significantly improves throughput compared to a non-coordinated system.  To reduce computational complexity, we also provide a heuristic solution to the problem. The proposed heuristic achieves throughput comparable to that of the computationally expensive optimization tool based solution approach. 
\end{abstract}

\begin{IEEEkeywords}
Multi-AP coordination, WiFi 8, interference, C-SR, resource unit, nonlinear integer programming, throughput.
\end{IEEEkeywords}

\section{Introduction} \label{section:introduction}

\IEEEPARstart{C}urrently multi-access point coordination (MAPC) is a crucial research focus in the area of wireless local area networks (WLANs)  that significantly enhances network performance by fostering cooperation and making access points (APs) more aware of interference. This approach enables better management of network traffic, increases user throughput and spectral efficiency, and reduces both latency and packet loss.  This feature was first introduced in the amendment of WiFi 7 (IEEE 802.11be) to support a multi-AP WiFi environment~\cite{deng2020ieee,garcia2021ieee,khorov2020current}. However, this feature is considered as a key prospect of WiFi’s upcoming version, WiFi 8 (IEEE 802.11bn Ultra High Reliable (UHR)) to ensure high speed, ultra-high reliable and low latency communications~\cite{galati2024will},  \cite{reshef2022future}. 
 
There are various approaches to AP coordination, such as coordinated spatial reuse (C-SR), coordinated time division multiple access (C-TDMA), coordinated orthogonal frequency division multiple access (C-OFDMA), coordinated beamforming (C-BF),  and joint transmission (JTX) \cite{verma2023survey}. Among them, in recent years, the research community has shown significant research interests in C-OFDMA, C-SR, and C-BF approaches. OFDMA was a pivotal feature introduced in IEEE 802.11ax, enabling efficient sharing of bandwidth among multiple stations (STAs)~\cite{khorov2018tutorial}. This technology allows concurrent uplink or downlink transmissions by dividing the frequency spectrum into smaller sub-channels called Resource Units (RUs). However, its application in dense network environments is limited due to co-channel interference by other closely positioned APs and inefficiencies in resource allocation. To address thes challenges, C-OFDMA has been introduced to optimize RU assignment and enhance network capacity through MAPC \cite{lacalle2022multi}, \cite{imputato2024meeting}. However, C-OFDMA does not reuse the RUs by adjusting the transmission power of the coordinating APs. On the other hand, SR was another key technique introduced in IEEE 802.11ax to enhance spectral efficiency by enabling simultaneous transmissions in dense WLAN environments \cite{wilhelmi2021spatial}. However, the traditional SR techniques rely on static Transmission Power Control (TPC) mechanisms, leading to suboptimal interference management. To overcome this, C-SR  employs dynamic and cooperative power control strategies among multiple APs, which ensures channel re-usability adopting the dynamic nature of the network \cite{verma2023survey}, \cite{imputato2024beyond}. C-BF  also re-uses the RUs more efficiently than  C-SR by creating beams in the different directions. However, the transmission overhead in C-BF is significantly high as the channel state information (CSI) is required to implement it instead of  received signal strength indicator (RSSI). Moreover, error in CSI estimation can significantly degrade the performance of C-BF. 

Several research studies have been conducted on efficient RU allocation in APs and STAs \cite{yang2021mac}, \cite{islam2022proportional}, 
\cite{kuran2020throughput}. In \cite{yang2021mac}, the authors discuss several spatial reuse methods and provide a mechanism for the allocation of RUs in different STAs of a WLAN. The authors of \cite{islam2022proportional} and \cite{kuran2020throughput} have taken different RU scheduling approaches where they have built a mechanism to maximize throughput. In \cite{islam2022proportional},  the authors optimally distribute RUs among STAs based on their available loads to improve fairness. Throughput-maximizing OFDMA scheduler, considering client load, modulation coding schemes, and fairness metrics to optimize throughput, is presented in \cite{kuran2020throughput}. On the other hand, C-SR have been studied in many research works~\cite{imputato2024beyond}, \cite{nunez2022txop, 10811972,10978433,11063419, 10168857}. In \cite{imputato2024beyond}, the authors built an interference model to implement C-SR where APs within the same group can transmit simultaneously in a single  Transmit Opportunity (TXOP). The authors in \cite{nunez2022txop}, investigate the implementation of TXOP sharing in multi-AP cooperative IEEE 802.11be WLANs, aiming to improve throughput and reduce contention using C-TDMA and C-TDMA/SR. In \cite{10811972}, the authors propose AP-station pairing using reinforcement learning approach called multi-armed bandits (MABs) and then C-SR group selection problem is solved in each TXOP for suitable simultaneous AP-STA pair transmission.  In \cite{10978433}, throughput is  modeled in term of packet loss, throughput maximization problem is formulated by state machine  and then multi-agent reinforcement learning-based approach is considered for C-SR. In \cite{11063419}, the authors solve two problems: the first problem is time allocation in different transmission sets and the second problem is C-SR in each transmission set for maximizing price using machine learning. In \cite{10168857}, the authors propose a methodology for AP grouping and group scheduling for C-SR.   

Optimal RU allocation in multi-AP scenarios depends on the C-SR of RUs. If both are performed independently, the optimal performance cannot be achieved. To the best of our knowledge, joint optimization problem of RU allocation and C-SR has not been taken for MAPC WiFI systems. This research focuses on finding the optimal throughput performance with joint RU allocation and C-SR design in one TXOP of a MAPC WiFi system. We consider a WiFi network where multiple APs can interfere with each other. The main contributions in this paper are as follows:
\begin{itemize}
        \item We formulate an optimization problem to maximize network throughput by optimal allocation of RUs and transmit power levels among the users with optimal AP grouping. 
        \item We solve the problem for several network scenarios with BARON  optimization tool. 
        \item We propose a heuristic solution approach for the optimization problem so that computation complexity is reduced. 
        \item We compare the performances of the optimal and heuristic solutions with that of a non-coordinated operation of  networks and  investigate the effect of various parameters on throughput performance. 
\end{itemize}

The rest of the paper is organized as follows, Section~\ref{section:system_model} describes system architecture, system operation, and problem formulation. In Section~\ref{section:solution_approach}, we discuss the solution approaches including heuristic solution approach and non-coordinated network operation as benchmark. In Section~\ref{section:results}, we present the numerical results. Finally,  Section~\ref{section:conclusion} concludes the letter.

\section{System Model and Problem Formulation} \label{section:system_model}

\subsection{Network Architecture and Operation}

\begin{figure} 
    \centering    \includegraphics[width=0.8\linewidth]{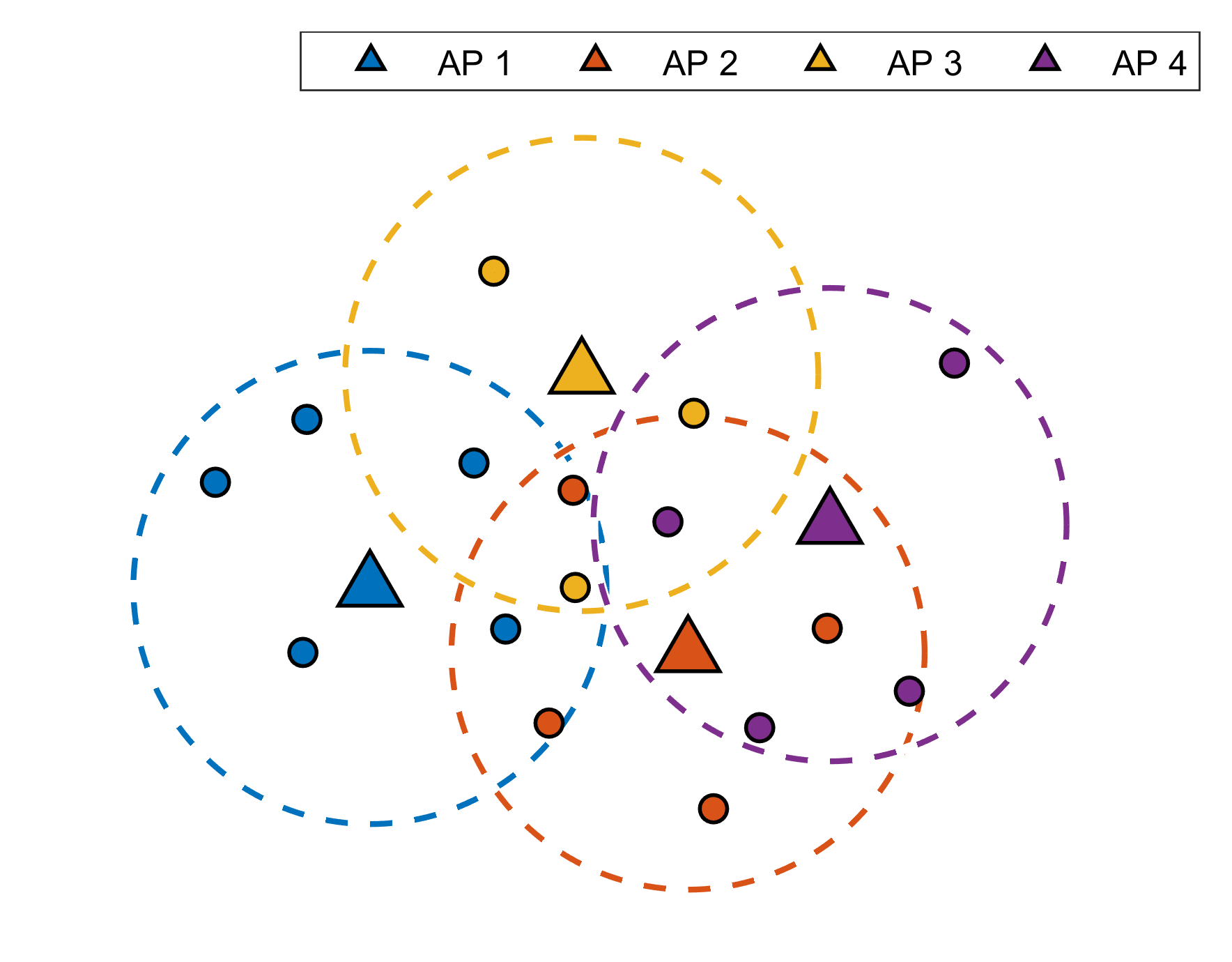}
    \caption{A typical MAPC WiFi system}
    \label{fig:system_model}
\end{figure}

We have considered a centralized MAPC (C-MAPC) based WiFi network with $N$ APs and $U_T$ users with overlapping basic service sets (OBSSs). A typical network diagram is shown in Fig.~\ref{fig:system_model}. The sets of the APs and the set of STAs are denoted by $\mathcal{N}$ and $\mathcal{U}_T$, respectively. The set of STAs associated to the AP $n\in\mathcal{N}$ is denoted by $\mathcal{U}_n$ and  $U_T=|\mathcal{U}_1|+|\mathcal{U}_1|+\ldots+|\mathcal{U}_N|$. The total number of RUs for the entire network is $J$ and their set is denoted by $\mathcal{J}$. The maximum number of AP groups is $G_{max}$ and the set of AP groups is denoted by $\mathcal{G}$. We assume that the APs of the same group will share the same set of RUs for their STAs. The maximum number of RUs per STA is assumed to be one.    The APs are connected to a master controller (MC) via a high-speed wired connection. The MC receives all RSSI information of the STAs through their associated APs, solves the optimization problem for AP grouping, RU allocation, and power assignment, and makes the scheduling decision. In this research, we only consider downlink transmissions. 

\subsection{Wireless Propagation Model}
We consider the log-distance path loss model  for wireless propagation \cite{log2022khaled}. The path loss between an AP and a user with distance $d$ is given by 
\begin{equation}
    PL(d) [dB] = PL(d_0)[dB] + 10\eta \log_{10} \left( \frac{d}{d_0} \right) 
    \label{eq:log_distance_path_loss_model}
\end{equation}
where, $\eta$ is the path loss exponent and $PL(d_0)$ is the path loss at a reference distance $d_0$. The path loss at reference distance $d_0$ is given by  $PL(d_0)[dB] = 20 \log_{10} \left( \frac{4 \pi d_0}{\lambda}\right)$ for an operating  wavelength $\lambda$. The channel gain between an AP and a user with distance $d$, $H(d)$ can be calculated as
\begin{equation}
    H(d) = 10^{-\frac{PL(d)[dB]}{10}}.
   \label{eq:channel_gain_formula}
\end{equation}

\subsection{Problem Formulation} \label{section:problem_formulation}
Let \( X_{u,j} \in \{0,1\} \) be a binary variable indicating whether RU \( j \) is assigned to user \( u \). We assume that \( P_{u,j} \) is the power allocated to user \( u \) on RU \( j \). Denote by $H_{u,AP[k]}$  the channel gain between the user \(u\) and the AP of user \(k\). Let the binary variable \(\lambda_{n,g} \in \{0,1\}\) indicates whether AP $n$ is in group $g$ or not and the binary variable $ y_g \in \{0,1\}$ indicates whether the group $g$ is active or not. Let  $\mathbf{X}$, $\mathbf{P}$, $\mathbf{y}$, and $\mathbf{\lambda}$ be the vectors for the $X_{u,j}$, $P_{u,j}$, $y_g$, and $\lambda_{n,g}$ variables.  The optimal AP grouping, RU allocation and power allocation problem with C-SR to maximize the total throughput of the network can be formulated as in (\ref{objective:max_throughput})-(\ref{constraint:active_group_indicator}). 
The optimization problem is a non-linear integer programming. In the objective function in \eqref{objective:max_throughput}, $SINR_{u,j}$ is the signal to interference plus noise ratio of a user $u \in \mathcal{U}_T$ for a RU $j \in \mathcal{J}$. Thus, the objective is to maximize the total throughput of the system.  $SINR_{u,j}$ is expressed by the constraints in \eqref{constraint:sinr_defination}, where $H_{u,AP[u]}$ is the channel gain between the user $u$ and the AP of the user $u$ and $N_0$ is the background noise power. The constraints in  \eqref{constraint:one_ru_per_user} ensure that an STA can be assigned a maximum of one RU. The constraints in \eqref{constraint:max_power_per_sta} limit the scheduler to allocate a maximum power of \( P_{\text{max}} \)  for any STA for a given RU. The constraints in \eqref{constraint:no_ru_reuse_within_ap} state that a RU will be used a maximum of one time within any BSS. The constraints in \eqref{constraint:max_power_per_ap} indicate that an AP  can use total \( P_{\text{max}}^{\text{AP}} \) power to transmit to all of its own users within its BSS.  The constraints \eqref{constraint:ap_group_assignment} prevent an AP from being a member of multiple groups simultaneously. The constraints in \eqref{constraint:max_ap_per_group} limit the number of APs in a particular group. The constraint in \eqref{constraint:group_cap} limits the number of active AP group in the system to $G_{max}$. The constraints in \eqref{constraint:group_based_ru_reuse} allows two users from different APs to share the same RU only if their APs are in the same group. The rest of the constraints are used to represent binary variables or to bound the power to zero value.  

\begin{align}
    \max_{\mathbf{X}, \mathbf{P}, \mathbf{y}, \mathbf{\lambda}}  \quad & \sum_{u \in \mathcal{U}_T}  \sum_{j \in \mathcal{J}} \log_2 (1+SINR_{u,j}) \label{objective:max_throughput}\\
    SINR_{u,j} &\leq 
    \frac{P_{u,j} X_{u,j} H_{u,AP[u]}}
         {\sum_{\substack{k = 1, k \neq u}}^{U_T} 
         P_{k,j} X_{k,j} H_{u,AP[k]} + N_0} \nonumber\\         
    &\qquad \qquad \qquad  \forall u \in \mathcal{U}_T,\ \forall j \in \mathcal{J}
    \label{constraint:sinr_defination}\\
    \sum_{j=1}^{J} X_{{u},j} &\leq 1 \quad \forall u \in \mathcal{U}_T \label{constraint:one_ru_per_user}\\
    P_{u,j} &\leq X_{u,j}  P_{\text{max}} \quad \forall u \in \mathcal{U}_T \ ,\ \forall j \in \mathcal{J}
    \label{constraint:max_power_per_sta}\\
    \sum_{{u} \in \mathcal{U}_n} X_{{u},j}& \leq 1 \quad \forall n \in \mathcal{N} \quad \forall j \in \mathcal{J} \label{constraint:no_ru_reuse_within_ap}\\
    \sum_{{u} \in \mathcal{U}_n} \sum_{j \in \mathcal{J}} P_{{u},j} &\leq P_{\text{max}}^{\text{AP}} \quad \forall n \in \mathcal{N} \label{constraint:max_power_per_ap}\\
    \sum_{g \in \mathcal{G}} \lambda_{n, g} &= 1 
    \quad \forall n \in \mathcal{N}
    \label{constraint:ap_group_assignment}\\
    \sum_{n \in \mathcal{N}} \lambda_{n, g} &\leq N y_g
    \quad \forall g \in \mathcal{G}
    \label{constraint:max_ap_per_group}\\
    \sum_{g \in \mathcal{G}} y_g &\leq G_{\max}
    \label{constraint:group_cap}\\
    X_{u j} + X_{v j} &\leq 1 + \sum_{g \in \mathcal{G}} 
    \lambda_{\mathrm{AP}[u],g} \cdot \lambda_{\mathrm{AP}[v],g} \nonumber \\
    &\quad \forall u,v \in \mathcal{U}_T, \ u<v, \ \forall j \in \mathcal{J}
       \label{constraint:group_based_ru_reuse}\\
  X_{u,j}   &\in \{0,1\} \quad \forall u \in \mathcal{U}_T \ , \ \forall j \in \mathcal{J}
    \label{constraint:ru_allocation_variable}\\
   P_{u,j}  &\geq 0 \quad \forall u \in \mathcal{U}_T \ , \ \forall j \in \mathcal{J}
    \label{constraint:power_allocation_variable}\\
    \lambda_{n,g} &\in \{0,1\} \quad \forall n \in \mathcal{N} \ , \ \forall g \in \mathcal{G}
    \label{constraint:ap_group_variable}\\
   y_g & \in \{0,1\} \quad \forall g \in \mathcal{G}    \label{constraint:active_group_indicator}
\end{align}

\section{Solution Approaches and Benchmark} \label{section:solution_approach}
\subsection{Solution by Optimization Tool} 
We have used a mathematical programming language (AMPL) to model the optimization problem, which is well-suited for large-scale optimization. The model is later solved using the BARON, which is a global optimization solver. The solver is accessed through NEOS server, which provides cloud-based computational resources for optimization.

\subsection{Proposed Heuristic Solution} 
The proposed heuristic algorithm allocates STAs and their transmit power levels for each RU in a sequential manner, as described in Algorithm \ref{algorithm:heuristic}. For a RU $j \in \mathcal{J}$, the algorithm first tries to allocate $G_j=\frac{{U}_T}{J}$ number of STAs considering all the possible STA-Power combinations with three power levels  $P_{min}$, $P_{mid}$, and $P_{max}$. To do this, first, the STAs are sorted according to the descending order of channel gain from their APs which is denoted by $\mathcal{U}_s$ and then, the set of STA sets with size $G_j$, $\mathcal{Y}$ are determined keeping the STA with highest channel gain, along with the STAs from other APs which are far away from it. For each set of STAs $y \in \mathcal{Y}$, STA-Power combination set $\mathcal{Z}_y$ is determined. To reduce the computation burden, the capacity of an STA-power combination is calculated only when the minimum SINR of all the STAs in $y$ is higher than a threshold $\gamma_{th}$. A STA-power combination which provides maximum capacity is  selected as STA-Power combination for the RU $j$ which is denoted as $D_j$. The set $\mathcal{U}_s$ is then updated by removing the STAs in $D_j$. Moreover, from all the assigned STA-power combinations, the power requirement of each AP is calculated and if the total power requirement in an AP is higher than $P_{max}^{AP}-P_{min}$, then all the STAs of the AP is removed from  $\mathcal{U}_s$. If $D_j$ is found empty for a STA size $G_j$,  the size $G_j$ is reduced by one, and similar procedure is applied to find $D_j$ for the RU $j$.  

\subsection{Benchmark} 
We have used the non-coordination allocation approach \cite{parizi2025ieee} as a benchmark for comparison. In this process, there will be no coordination among the APs. Each AP  independently allocates RUs to its corresponding STAs. RUs are allocated randomly, and the downlink transmit powers are set to the maximum value.  

\begin{algorithm}[t]
\caption{Heuristic Algorithm}
\label{algorithm:heuristic}
\begin{algorithmic}[1]
\REQUIRE $N$, $U_T$, $J$, $P_{min}$, $P_{mid}$, $P_{max}$, $\mathbf{H}$, $N_0$, $\gamma_{th}$
\ENSURE For each RU $j \in \mathcal{J}$, set of users and their transmit power levels, $D_j $
\STATE Generate set of STAs according to decending order of channel gain, $\mathcal{U}_s$ 
\FOR{$j = 1$ to $J$}
\STATE $G_j =  U_T/J, D_j= \{\phi\}, C_0=0$
    \WHILE{$\mathcal{U}_s \neq \{\phi\}$}
        \STATE Select the first STA from $\mathcal{U}_s$
        \STATE Sort the APs according to low interference order to the first STA
        \STATE Form set of STA sets each with size $G_j$, $\mathcal{Y}$ by including the first STA and  STAs from the sorted APs one after another 
        \FOR{Each $y\in\mathcal{Y}$}
        \STATE Create sets of all possible STA-power combination, $\mathcal{Z}_y$ for STA set $y$
            \FOR{Each $z \in \mathcal{Z}_y$} 
                \STATE Calculate min SINR of the STAs in $y$, $S_{min}$
                \IF{$S_{min} \ge \gamma_{th}$}
                    \STATE Calculate capacity, $C_z=\log_2(1+S_{min})$
                    \IF{$C_z \ge C_0$}
                        \STATE$D_j=z$ \& $ C_0=C_z$
                    \ENDIF
                \ENDIF
            \ENDFOR
        \ENDFOR
        \IF{$D_j \neq \{\phi\} $}
               \STATE Remove STAs of $D_j$ from 
               $\mathcal{U}_s$. For each AP, calculate total power ($P_{total}$) for the STAs in $D_j, j\in \mathcal{J}$  and remove all STAs of an AP $n\in\mathcal{N}$ from $\mathcal{U}_s$ if $P_{max}^{AP}-P_{total}\le P_{min}$ for the AP $n$ 
            \STATE \textbf{break} 
        \ELSE
            \STATE Reduce conflict-free set size: $G_j = G_j - 1$
            \IF{$G_j = 0$}
                \STATE \textbf{break} 
            \ENDIF
        \ENDIF
    \ENDWHILE
\ENDFOR

\RETURN $D_j, j\in \mathcal{J}$
\end{algorithmic}
\end{algorithm}

\section{Results} \label{section:results}

\begin{table}
\centering
\caption{The values of the different parameters}
\label{table:simulation_parameters}
\begin{tabular}{|c|l|c|}
\hline
\textbf{Parameter} & \textbf{Notation Description} & \textbf{Value} \\ \hline
$f$ & Frequency Band & 2.4 GHz \\ \hline
$P_{\max}$   & Max Power for each STA & 15 mW \\ \hline
$P_{max}^{AP}$  & Max Power of each AP & 100 mW \\ \hline
$\eta$ & Path Loss Exponent & 2.5 \\ \hline
$d_0$ & Reference distance & 1 m \\ \hline
$N_0$ & Noise Power & -96 dBm \\ \hline
$G_{max}$ & Max. No. of group & 4 \\ \hline
$\gamma_{th}$ & SINR Threshold & 2 dB\\ \hline
\end{tabular}
\label{tab:sim_params}
\end{table}

We take four fixed-coordinate APs form overlapping BSSs in a 2D plane. The STAs are randomly assigned so each AP has at least one STA. Each AP has coverage radius \(R = 10\) m and average inter-AP distance 11.74 m. The locations of the STAs are determined using polar coordinates \((r, \theta)\) with a clustering model where a certain percentage users are near to the AP with \(r\) Rayleigh-distribution of parameter $\sigma=\frac{R}{a}$ and the rest of the STAs have \(r\) uniformly distributed in \([0.5R, R]\) to represent edge/overlap users. Five instances are generated by varying the percentage 30-60\% with five values of $a=\{2, 2.5, 3 ,3.5, 4\}$. For all the STAs, \(\theta \sim \text{Uniform}[0, 2\pi)\). Note the five network instances are considered for a given number of STAs for averaging the performance. The total number of STAs of the system varies from 8 to 24. We consider the primary channel with a bandwidth of 20 MHz and 10 RUs, each of which is 2 MHz. The rest of the 
parameters used for performance evaluation are presented in Table \ref{table:simulation_parameters}. The RSSIs of the STAs are determined using the channel model in (\ref{eq:channel_gain_formula}). The values of $P_{min}$ and $P_{mid}$ are taken to be 5 mW and 10 mW, respectively for the heuristic algorithm. 


\begin{figure}
\centering
\begin{minipage}{0.35\textwidth}
    \centering
    \includegraphics[width=\textwidth]{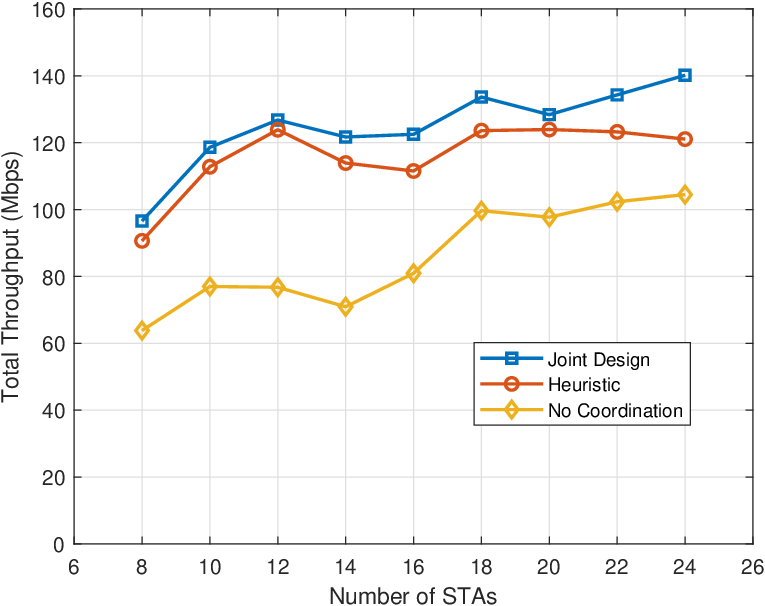}
    \caption{Total throughput of the system  for different number of STAs.}
    \label{fig:throughput_vs_stas}
\end{minipage} 

\vspace{1em} 

\begin{minipage}{0.35\textwidth}
    \centering
    \includegraphics[width=\textwidth]{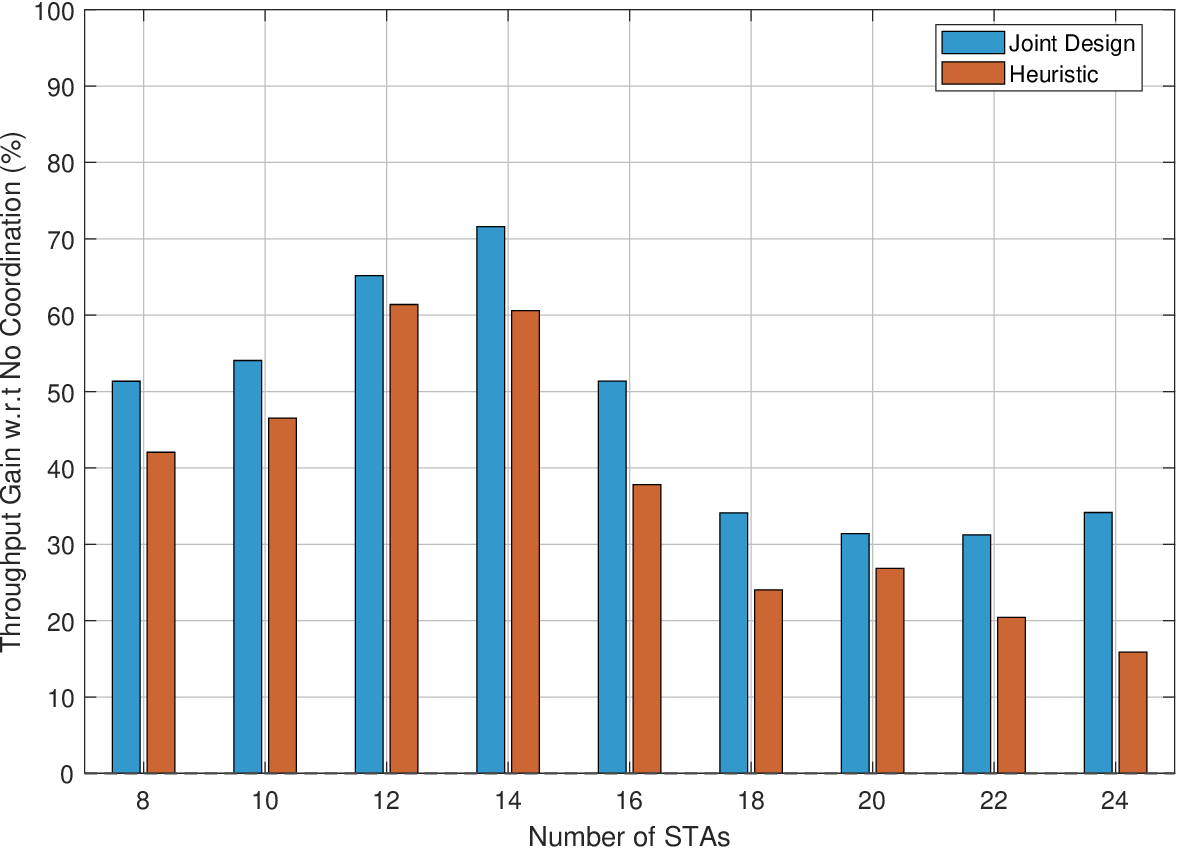}
    \caption{Throughput gains for different number of STAs in the system.}
    \label{fig:throughput_gain_vs_users}
\end{minipage}
\end{figure}

\begin{figure}
\centering
\begin{minipage}{0.35\textwidth}
    \centering
    \includegraphics[width=\textwidth]{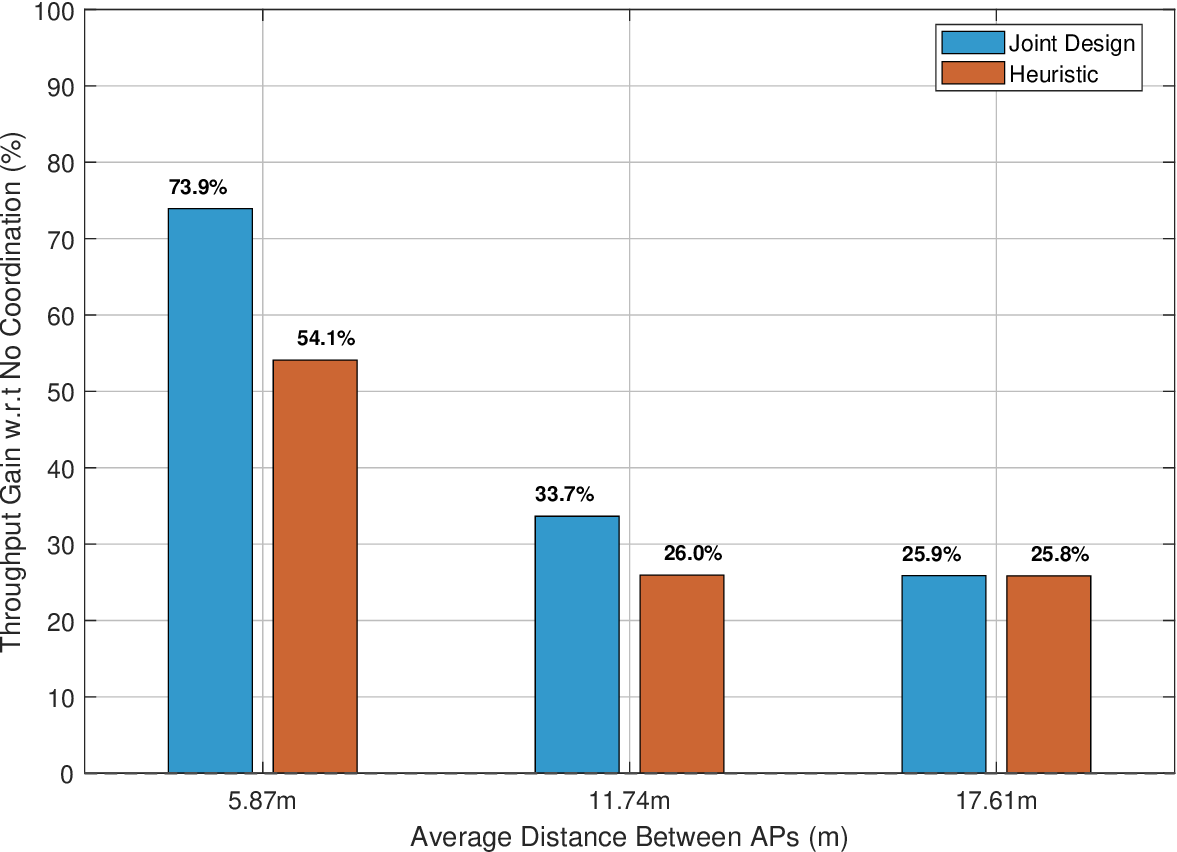}
    \caption{Effect of inter AP distance on throughput gains.}
    \label{fig:throughput_gain_vs_distance}
\end{minipage} 

\vspace{1em} 
\begin{minipage}{0.35\textwidth}
    \centering
    \includegraphics[width=\textwidth]{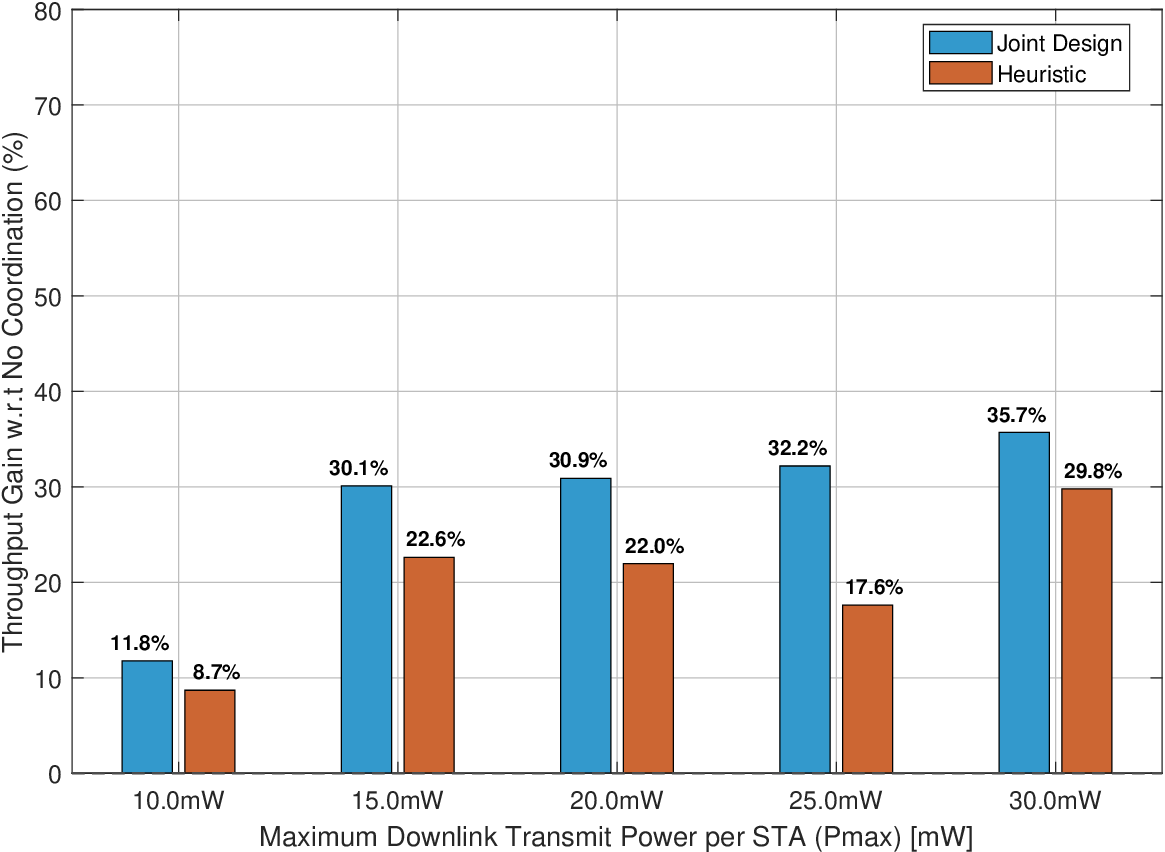}
    \caption{Effect of the value of maximum transmit power on throughput gains.}
    \label{fig:throughput_gain_vs_transmit_power}
\end{minipage}
\end{figure}

We present the total throughput of all STAs in the system in Fig. \ref{fig:throughput_vs_stas} while the number of STAs is varied from 8 to 24. The results show that the joint approach achieves significantly higher throughput than that of the non-coordinated system. The heuristic model also outperformed the non-coordination scenario. In Fig.\ref{fig:throughput_gain_vs_users}, we show the throughput gains of optimization tool based solution approach and the heuristic based solution approach with respect to the non-coordinated system. Both the solution approaches show significant throughput gains. When the number of STAs is lower than the number of RUs, the throughput gain is lower and it increases with increasing the number of STAs as spatial reuse increases. When the system has 14 STAs in total, the gain is found to be the  highest and it is around 71.6\% and 70.05\% for the joint design and heuristic design, respectively.  However, throughput gains start to decrease for further increment of the number of users. It is  due to that fact that the interference to the C-SR users increases with increasing the number of users, as a higher number of users have to use the same RU. When the number of users is 24, the gains reduce to 34.2\% and 15.9\% for the joint design and the heuristic design, respectively. 

In Fig.\ref{fig:throughput_gain_vs_distance},  throughput gains of joint and heuristic designs are compared to non-coordinated allocation for different intra-AP distances with $U_T=16$, keeping the same relative distance of the users from the APs. We find that when the APs are closer, the benefit of the joint design is the higher and the benefit reduces with increasing the distance between the APs. Throughputs at 5.87m, 11.74m, and 17.61m distances for the joint design are found to be 129.3 Mbps, 132.9 Mbps, and 143 Mbps, respectively compared to 74.3 Mbps, 99.5 Mbps, and 113.6 Mbps throughputs under the non-coordination system. We attribute this to the fact that interference to the users is higher when the APs become closer and the system throughput becomes lower. However, the non-coordinated system suffers from interference significantly, but the joint system manages the interference effectively. When the distance between the APs increases, user-interference decreases, and the gains from the joint design decrease, while the system's overall throughput increases. 

In Fig. \ref{fig:throughput_gain_vs_transmit_power}, we present the effect of maximum transmit power to each of the users on the throughput gains of the joint design and the heuristic design with $U_T=20$. The joint design achieves throughputs of 121.96 Mbps, 132.9 Mbps, 139.28 Mbps, 137.85 Mbps, 145.2 Mbps at 10 mW, 15 mW, 20 mW, 25 mW, and 30 mW transmission power, respectively. The results show that throughput and throughput gains increase with increasing limit of maximum transmit power to each user. We attribute this to the fact that if the limit of transmit power is higher compared to the noise power,  the flexibility of interference management  is greater. From (\ref{constraint:sinr_defination}), it can be stated that this flexibility will become static when the transmit power is significantly high compared to the noise power and then the gain will not change even when the transmit power is further increased.

\section{Conclusion} \label{section:conclusion}
We have investigated the throughput gain achieved through joint RU allocation and C-SR in coordinated multi-AP WiFi system by formulating and solving an optimization problem. Due to complexity, the optimal solution approach is not practical. However, this study highlighted the importance of joint design in improving throughput performance. Furthermore, we provided a simple heuristic solution that achieves performance close to the optimal solution. Considerable future work remains at the protocol, hardware, and algorithmic levels to develop more efficient heuristics or machine learning based solutions and enable practical real-world deployment.

\bibliographystyle{IEEEtran}
\bibliography{references.bib}

\end{document}